\documentclass[prl,twocolumn,superscriptaddress,showpacs,floatfix]{revtex4-1}
\usepackage{epsfig}
\usepackage{color}
\usepackage{graphics}
\begin{document}

\title{A Fragile-Strong Fluid Crossover and Universal Relaxation Times in a Confined Hard Disc Fluid.}

\author{Mahdi Zaeifi Yamchi}
\affiliation{Department of Chemistry, University of Saskatchewan,
Saskatoon, Saskatchewan, S7N 5C9}

\author{S. S. Ashwin}
\affiliation{Department of Chemistry, University of Saskatchewan,
Saskatoon, Saskatchewan, S7N 5C9}
\affiliation{Department of Mechanical Engineering and Material Science \&
Department of Physics,\\
Yale University, New Haven, CT 06511}

\author{Richard K. Bowles}\thanks{Corresponding Author: richard.bowles@usask.ca}
\affiliation{Department of Chemistry, University of Saskatchewan,
Saskatoon, Saskatchewan, S7N 5C9}
\email{richard.bowles@usask.ca}

\date{\today}

\begin{abstract}
We show that a system of hard discs confined to a narrow channel exhibits a fragile-strong fluid crossover located at the maximum of the isobaric heat capacity and that the relaxation times for different channel widths fall onto a single master curve when rescaled by the relaxation times and temperatures of the crossover. Calculations of the configurational entropy and the inherent structure equation of state find that the crossover is related to properties of the jamming landscape for the model but that the Adam-Gibbs relation does not predict the relaxation  behavior. We also show that a facilitated dynamics description of the system, where kinetically excited regions are identified with local packing arrangements of the discs, successfully describes the fragile-strong crossover.
\end{abstract}

\maketitle

Upon cooling or compression, many materials, including supercooled liquids\cite{Debenedetti:2001p1730,Debenedetti:1996tf}, gels and polymers, form amorphous, glassy solids, where the time needed for the system to structurally rearrange becomes longer than the experimental measurement. However, despite the ubiquitous appearance of glasses in nature and science, a comprehensive understanding of this glass transition remains elusive. One useful approach has been to classify glass forming liquids on the basis of the temperature dependence of their structural relaxation times, $\tau$, on approach to the glass transition temperature $T_g$ \cite{AustenAngell:1995p415}. In a strong liquid, $\tau$ is linear in an Arrhenius plot of $\ln\tau$ vs $1/T$, implying structural relaxation is a simple activated process. Fragile liquids exhibit super-Arrhenius behavior where $\tau$ increases much faster and appears to diverge at positive temperatures. This suggests relaxation is cooperative and has led to speculation that there is a thermodynamic ideal glass transition underlying the kinetic behavior observed in experiment and simulation~\cite{Debenedetti:2001p1730,Angell:1997uo}.

Silica~\cite{SaikaVoivod:2001p5554,SaikaVoivod:2004p10718}, silicon~\cite{Sastry:2003p14289} and water~\cite{Stanley:2008p1411,Angell:2008bx,Poole:2011bb,Xu:2005p14274} all appear to go through a dynamical crossover from a fragile liquid to a strong liquid at a crossover temperature, $T_{\times}$, that coincides with the Widom line, marked by a maximum in the heat capacity. However, a recent analysis of the transport coefficients of 84 different glass formers shows that the fragile-strong (FS) crossover occurs more widely than originally thought and suggests that the crossover temperature may be more relevant to the general features of dynamical arrest~\cite{Mallamace:2010p14279} than $T_g$, which is based on an arbitrarily chosen experimental relaxation time.

In this letter, we study two contrasting  paradigms used to describe glassy  behavior: The first is the inherent structure landscape (ISL)~\cite{Stillinger:1964jp,Stillinger:1982jb}, or the density packing landscape, which is the hard particle equivalent to the potential energy landscape~\cite{Goldstein:1969uq}, and the second is the facilitated dynamics~\cite{Fredrickson:1984co,Garrahan:2003jq} (FD) approach. Both approaches are shown to capture key elements of the crossover but we also find that the local packing arrangements of discs, which ultimately give rise to the ISL, can be related to the kinetically excited regions appearing in the FD paradigm, providing a connection between the two approaches.





Our model consists of $N$ two-dimensional (2D) hard discs, with diameter $\sigma$,  confined between two hard walls (lines) of length $L$ separated by a distance $1< H_d/\sigma < 1+\sqrt{3/4}$. The particle-particle and particle-wall interaction potentials are given by
\begin{equation}
\begin{array}{lcl}
V_{r_{ij}}= \left\{\begin{array}{ll}
0 & r_{ij} \geq \sigma \\
\infty & r_{ij} <  \sigma
\end{array}
\right .
&:&
V_{w}(r_{i})= \left\{\begin{array}{ll}
0 & r_{y} \leq  \left | h_0/2 \right | \\
\infty & \textup{otherwise}
\end{array}
\right.
\end{array}
\label{eq:eqn1}
\end{equation}
respectively, where $r_{ij}=\left | \mathbf{r_{j}-r_{i}} \right |$ is the distance between particles,  $r_{y}$ is the component of the position vector for a particle perpendicular to the wall and $h_0=H_d-\sigma$. The occupied volume is $\phi=N\pi \sigma ^{2}/\left(4LH_d\right)$. 

\begin{figure}[t]
\includegraphics[width=2.5in]{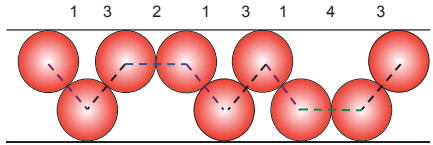}
\caption{Local packing arrangements that can be combined to give collectively jammed states. Dashed lines connect the centers of neighboring discs in contact and the numbers identify different ``bonds".  Bonds 1 and 3 are the locally most dense states. Bonds 2 and 4 represent the defect states.}
\label{fig:pack}
\end{figure}

The exact partition function~\cite{KOFKE:1993p3863} and complete jamming phase diagram~\cite{Bowles:2006bs,Bowles:2011jh} for the system are known, making it an ideal tool for exploring the relationships between thermodynamics, dynamics and the ISL. Fig.~\ref{fig:pack} shows the four locally jammed packing configurations of the discs that can be combined to give the collectively jammed~\cite{Torquato:2001ts} inherent structures. The occupied volume fraction of the jammed states is  $\phi_{J} = \pi / \left\{ 4H_d \left[  \theta +  \left(1- \theta \right) \sqrt{\left(2-H_d \right) H_d} \right] \right \} $, where $\theta$ is the mole fraction of defects (type 2 and 4 bonds). The most dense jammed state, $\phi_{Jmax}$, occurs when $\theta=0$. The least dense jammed packing occurs when $\theta=0.5$ and consists of a repeating unit cell of $-1-4-3-2-$ bonds because placing two defect bonds together (-2-2-, or -4-4-) results in an unstable configuration. Recent studies of this model found that 
configurations of the ideal gas mapped to jammed states, $\phi_{Jig}$, corresponding to the maximum in the packing distribution with $\theta_{ig}=1/2-5^{1/2}/10$. The equilibrium fluid then only samples basins with a higher $\phi_J$ as $\phi$ is increased~\cite{Bowles:2006bs,bowles:prep}.

\begin{figure}[t]
\includegraphics[width=3.5in]{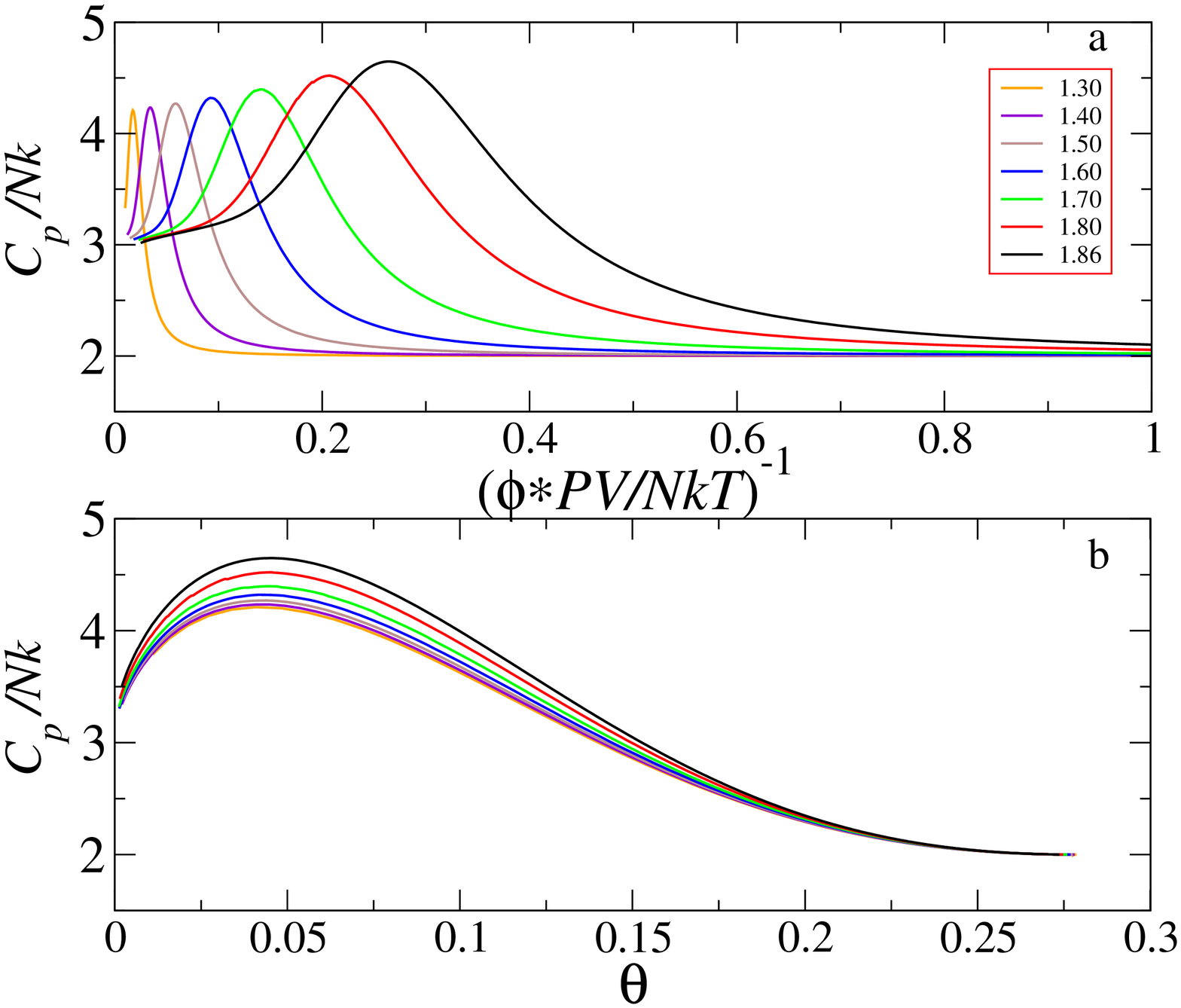}
\caption{$C_p/Nk$ for different values of $H_d$ as a function of (a) $(\phi PV/NkT)^{-1}$ and (b) $\theta$.}
\label{fig:cpmax}
\end{figure}

We obtain the exact equation of state (EOS) using the transfer matrix method developed by Kofke {\it et al}~\cite{KOFKE:1993p3863}.  If the positions of the discs are fixed in the $y$-direction, the configurational integral in the $x$-direction can be treated as a 1D mixture of hard rods on a line. Taking the Laplace transform, gives the partition function in the $N,P,T$ ensemble as
\begin{equation}
\Delta(N,P,T) = \frac{1}{\Lambda^{DN}(\beta P)^{N+1}} \int dy K^{N}(y,y)\mbox{. }\\
\label{eq:z1}
\end{equation}
Here, $\Lambda$ is the thermal wavelength, $P$ is the longitudinal pressure and $K(y_1,y_2)=\exp[-P h_0 L_x(y_1,y_2)]$, with $y_1$ and $y_2$ being the $y$-coordinates of two adjacent discs in contact. $L_x$ is the projection of the distance between the two contacting discs along the $x$-axis and is a function of $y_1,y_2$ and $h_0$. Solving the eigenvalue problem associated with Eq. \ref{eq:z1} yields the equation of state for the fluid in the thermodynamic limit. This gives us access to the heat capacity given by $C_{p} /Nk = \left(\partial H / \partial T \right)_{P}=1 + Z/\left( 1+ d \ln \left \{ Z \right \} /  d \ln \left \{ \phi \right \} \right )$, where $H=NkT+PV$ is the enthalpy in the hard discs system, $Z=PV/NkT$ is the compressibility and $V=h_0L$ is the volume accessible to the disc centers. Fig.~\ref{fig:cpmax}(a) shows that $C_p/Nk$ as a function of $T$ has a maximum. However, the $C_p$ maximum in this system results from the binomial density of states and is not connected to the presence of an underlying critical point associated with a liquid-liquid phase transition.


Adam and Gibbs~\cite{Adam:2004p11431} argued that the rapid slow down in the dynamics of fragile liquids results from the decrease in accessible configurations at low temperatures or high densities and predicted relaxation times to behave as, $\tau =A \exp \left (B/TS_{c} \right )$, where $A$ and $B$ are effectively constant, $S_c=k\ln(N_J)$ is the configurational entropy and $N_J$ is the number of inherent structure basins accessible to the equilibrium fluid. The Adam-Gibbs relation predicts a divergence in $\tau$ as $N_J\rightarrow 1$, causing $S_c\rightarrow 0$, and it has been used to describe the relaxation in a wide variety of materials\cite{Speedy:2001p5140,Sastry:2001p10710,Stanley:2000cw}.

\begin{figure}[t]
\includegraphics[width=3.5in]{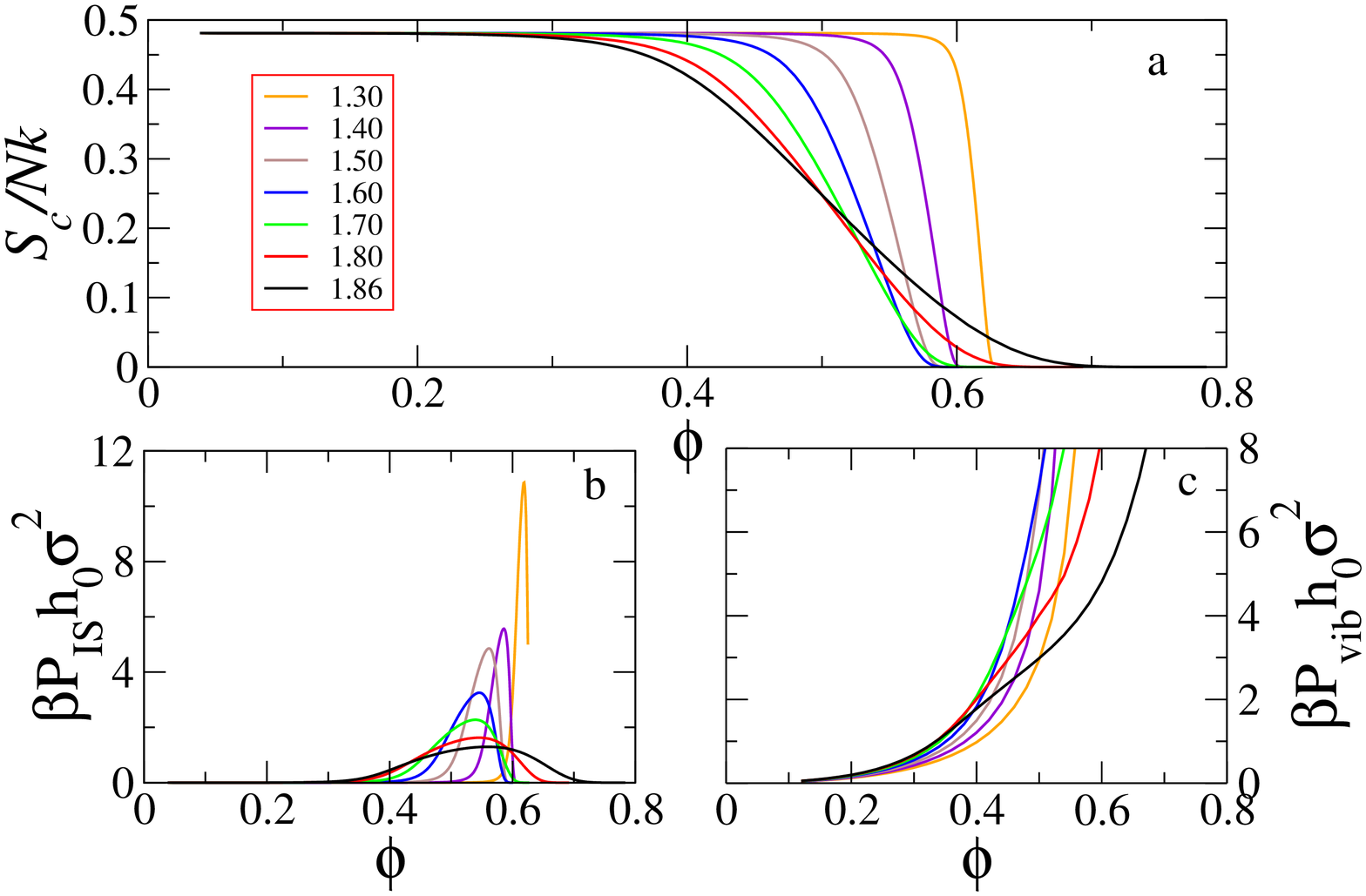}
\caption{(a) The equilibrium $S_c$ (b) $P_{IS}$ and (c) $P_{vib}$ as a function of $\phi$ for different values of $H_d$.}
\label{fig:sc}
\end{figure}

In our model, the configurational entropy~\cite{Bowles:2006bs} is given by $S_c(\phi)/Nk=(1-\theta)\ln(1-\theta)-\theta\ln\theta-(1-2\theta)\ln(1-2\theta)$,
where the equilibrium value of $\theta(\phi)$ can be obtained by analytically quenching the fluid to its local inherent structure using the transfer matrix method and the information about the local geometry of four discs in contact contained in the chain product matrix $K(y_i,y_m)K(y_m,y_n)K(y_n,y_j)$~\cite{bowles:prep}. Fig.~\ref{fig:sc}(a) plots $S_c$ as a function of $\phi$ and shows that the rate of configurational entropy loss increases with increasing $\phi$ at low densities but the impending Kauzmann catastrophe, thought to occur in fragile liquids, is avoided when $S_c$ goes through an inflection point. Consequently, the fluid has no ideal glass transition and $S_c$ only goes to zero in the limit $\phi \rightarrow \phi_{Jmax}$ and $PV/NkT\rightarrow\infty$. This is the expected behavior for a system where the distribution of packings is determined by localized point defects~\cite{Stillinger:1988es}. In addition, Fig.~\ref{fig:cpmax} (b) shows that the $C_p$ maxima, for the different $H_d$, all occur at the same $\theta=0.044\pm0.002$, which suggests it is the number of defects that controls the thermodynamics associated with the $C_p$ maximum.

To calculate the relaxation times for the system as a function of $\phi$, we use molecular dynamics (MD) simulations where the time, $t$, has units of $\sigma(m/kT)^{1/2}$. At any time during a MD trajectory, an instantaneous configuration can be mapped to its local inherent structure by considering the position of each disc relative to its two neighbors and using the triangular constraint, involving three neighboring discs, introduced by Speedy\cite{Speedy:2001p5140}. If the central disc lies below the line joining the centers of the outside discs, the central disc will pack at the bottom of the channel, otherwise it will pack at the top. This is equivalent to the analytical quench used to obtain $S_c$. Once the local packing positions have been identified, the bonds between neighboring discs are assigned their labels, 1-4 (see Fig.~\ref{fig:pack}). The fluid remains within the basin of a single inherent structure for a short time before a local rearrangement of the discs, which changes the identity of some of the bonds, moves the system to a new inherent structure.  We measure $R(t)$, the fraction of bonds that have not changed at least once in time $t$ as a function of $t$, and define the relaxation time as $\tau=\int_0^{\infty}R(t)dt$.

Our simulations were performed using $N=2000$ discs and periodic boundaries in the longitudinal direction. At each $\phi$ studied, $400N$ collisions were used to establish equilibrium after the system had been compressed from the previous $\phi$ using a modified version of the Lubachevsky and Stillinger algorithm~\cite{LUBACHEVSKY:1990p2185} that maintains a fixed ratio of $H_d/\sigma$. Simulation lengths varied from $200N$ collisions at low densities up to $10^6N$ collisions at high densities and 80000 configurations were mapped to their inherent structure at each $\phi$. $R(t)$ always decays to zero in the time scale of the measurement, suggesting the system remains a fluid for all densities considered.

\begin{figure}[t]
\includegraphics[width=3.5in]{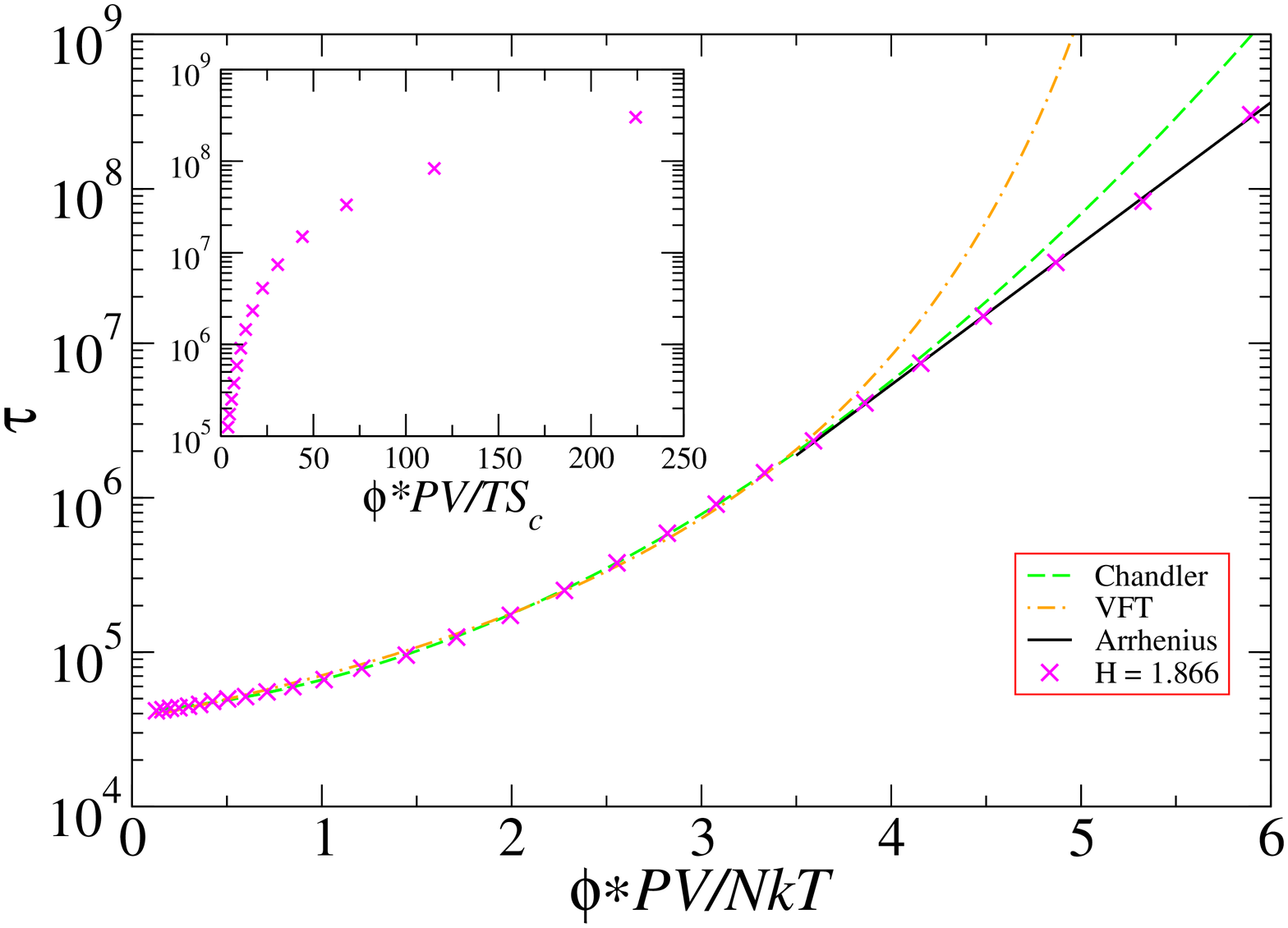}
\caption{Arrhenius plot for the relaxation times for $H=1.866$. The dashed and dashed-dot lines represent fits to the data in the fragile region of the parabolic and VFT equations respectively. The solid line is the Arrhenius fit to the strong fluid region. Insert: The Adam-Gibbs plot for $\tau$.}
\label{fig:fscross}
\end{figure}

For a hard particle system, $\phi PV$ is a constant along an isobar and the Arrhenius law would predict that $\ln\tau$ varies linearly with $\phi PV/NkT$. Fig.~\ref{fig:fscross} shows that $\tau$ increases more rapidly than the Arrhenius law predicts at low densities (high $T$), which suggests the fluid is fragile, but we see a crossover to strong-fluid behavior at high densities, where the relaxation times increase linearly. We also show fits of the data from the fragile region to the Vogel-Fulcher-Tammann~\cite{vogel.1921,Fulcher:1925fk,Tammann:1926uq}(VFT) equation, $\tau= A\exp \left [ B/(T-T_{VFT}) \right]$, which predicts a divergence of the relaxation times at a temperature $T_{VFT}>0$K, along with the parabolic law developed by Elmatad, Chandler and Garrahan~\cite{Elmatad:2009kx,Elmatad:2010hg}, which predicts no singularity and is derived on the basis of the FD models. Both equations fit well when restricted to fragile fluid data and the Arrhenius equation provides the best fit for $\tau$ above the crossover. Good fits of the VFT equation to a wide  range of experimental and simulation data for supercooled liquids have been used as evidence for the presence of a thermodynamic singularity underlying the experimentally observed glass transition. However, we have already shown that our model does not exhibit an ideal glass transition, which suggests  $T_{VFT}$ is simply a fit parameter with no physical significance. According to the Adam-Gibbs relation, $\ln \tau$ should vary linearly with $\phi PV/TS_c$, but the inset to Fig.~\ref{fig:fscross} shows that this is not the case here. Sastry et al~\cite{Sengupta:2012jg} recently found that the Adam-Gibbs relation did not hold in two-dimensions.

\begin{figure}[t]
\includegraphics[width=3.5in]{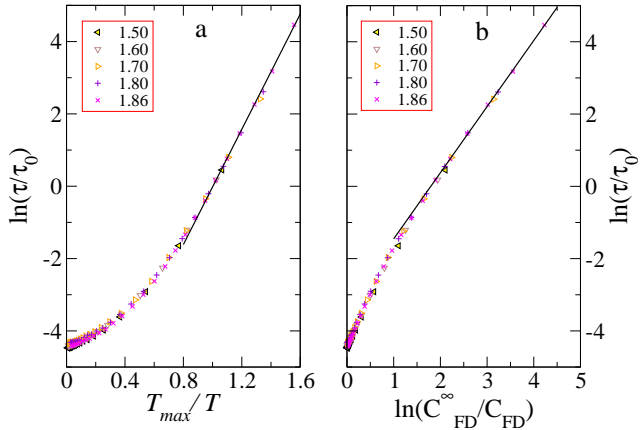}
\caption{Master curve for the $C_p$ maximum rescaling of the relaxation times and temperatures for the Arrhenius plot (a) and the facilitated dynamics model (b). The solid lines represent linear fits to the strong fluid region.}
\label{fig:scaling}
\end{figure}

The FS crossover occurs at the same $\phi$ as the maximum in the $C_p$ for all channel diameters, which is consistent with the studies that connect the crossover to the thermodynamics of the Widom line. Furthermore, for each $H_d$ we locate the temperature of the $C_p$ maximum, $T_{max}$, using our thermodynamic analysis, and define $\tau_0$ as the relaxation time at $T_{max}$. The temperatures and relaxation times are then rescaled by $T_{max}$ and $\tau_0$ respectively to give rise to the plot in Fig.~\ref{fig:scaling}(a), where all the curves have collapsed onto a single master curve. Rescaling by any other temperature, for example, by defining an arbitrary $T_g$ time scale, fails to collapse the data and leads to the impression that the systems with different $H_d$ have different fragilities. This suggests $T_{max}=T_{\times}$ provides a more useful scaling temperature for our model.

The relationship between the fragility of a fluid and the ISL has been explored in terms of the number and distribution of inherent structures~\cite{Speedy:1999ux,Sastry:2001p10710} but we can compare fragile and strong  behaviors within the same model. In particular, we calculate the equivalent of the inherent structure pressure~\cite{Shell:2003p4553} for a hard particle system directly from the configurational entropy as $P_{IS}=T(\partial S_c/\partial L)_{U,h_0}$, then use the relation $P=P_{IS}+P_{vib}$ and the exact EOS to isolate the vibrational contribution to the pressure (see Figs.~\ref{fig:sc}(b) and \ref{fig:sc}(c)). $P_{IS}$ makes a significant contribution to the overall pressure in the fragile fluid and reflects the fact that the configurational entropy of the system is varying rapidly as a function of density in the fragile fluid. However, $P_{IS}$ then goes through a maximum at a density slightly higher than that of the $C_p$ maximum and rapidly decreases. The EOS of the strong fluid is entirely dominated by the vibrational component of the system rattling around the local jamming point of the inherent structures being sampled. This is consistent with experimental findings\cite{Casalini:2005fq} that show strong behavior is dominated by density effects and local jamming while fragile relaxation is more thermally activated.

The facilitated dynamics paradigm suggests relaxation and particle motion is driven by local microscopic dynamical rules rather than any underlying thermodynamics~\cite{Garrahan:2003jq}. A key ingredient of FD is the presence of kinetically mobile regions that are able to influence the motion of neighboring regions, leading to the formation of chains of mobile particles in space-time. In addition, the theory argues that directed particle motion plays an important role. If a kinetically mobile region can activate or deactivate a neighboring region independent of any previous motion, it is considered to be directionally independent. Then the system behaves like a strong fluid and $\ln \tau\approx -\ln C_{FD}$, where $C_{FD}$ is the concentration of kinetically excited regions. The expectation that structural relaxation in a fragile fluid is cooperative is captured by having directional correlation between the successive movement of particles in the kinetically excited regions. A FS crossover is predicted to occur when elements of both mechanisms are present in the system.  While FD models have been parameterized to fit experimental data, only recently have there been efforts to identify the kinetically excited regions at a microscopic level~\cite{Keys:2011er,Sussman:2012gj,Jacquin:2012tb} and most studies of FD have focused on spin models where the dynamic rules are included by construction. 
 
 In the current model, we are able to identify these kinetically excited regions as the defects states (2, 4 bonds) in the jammed configurations. Relaxation in the system occurs in three ways: A) A particle next to an isolated defect can hop into the defect, causing the defect to move. This occurs with equal probability in both directions and leads to strong fluid behavior. B) Two neighboring defects moving toward each other create a local configuration with bonds $-1-4-4-3-$ or $-3-2-2-1-$ that is unjammed, leading to a spontaneous collapse of the central disc and an annihilation of the defects to form a $-1-3-1-3-$ locally jammed state. It is the spontaneity of the particle rearrangements associated with the defect annihilation, following the initial particle hop that brings the defects together, that is characteristic of the cooperative relaxation in a fragile fluid. C) A non-defect state can create two neighboring defects that move apart. The equilibrium number of kinetically excited regions in the system is then just  $C_{FD}=\theta(\phi)$, which we obtained from our analytical quench of the system. At low densities, there is a high concentration of defects that can interact and the directed creation-annihilation mechanisms dominate, giving rise to a fragile fluid that crosses over to a strong fluid as $C_{FD}$ decreases below its critical value.  Fig.~\ref{fig:scaling}(b) shows that $\ln \tau/\tau_0$ becomes a linear function of $-\ln (C^{\infty}_{FD}/C_{FD})$, where  $C^{\infty}_{FD}$ is the number of defects in inherent structures of the ideal gas, $\theta_{ig}$, highlighting the strong fluid behavior at low $C_{FD}$ (high densities). The figure also shows that the relaxation times for this model collapse to a single master curve when rescaled by $\tau_0$, which, on the basis of Fig.~\ref{fig:cpmax}(b), is equivalent to rescaling by the relaxation time of the system containing the critical number of defects associated with the crossover.

Our work suggests that the local packing environments of particles, along with the way they interact, may serve as the important microscopic ingredients in the FD paradigm and points to a new analysis that can be explored in higher dimensions. In the case of hard particles, it may be useful to identify local packing structures, or local tilings, in the jammed inherent structures~\cite{Ashwin:2009p7481} as defects. In more complex liquids, such as silica and water, the local geometry of a particle is dominated by the formation of a random tetrahedral network (RTN) of bonds. A recent study of the dynamics of ST2 water~\cite{Poole:2011bb}, both above and below $T_{\times}$, showed that the temperature dependence of the diffusion coefficient could be explained in terms of the concentration of local defects in the RTN. Similar results have been found in models of colloids~\cite{DeMichele:2006wl} and nanoparticle systems~\cite{Starr:2006gn} with highly directional, tetrahedral bonding. It is still not known if the movement of particles in relation to the RTN defects can be described by the FD model, but these studies, along with ours, strongly suggest local packing and particle geometries may play an important role in the dynamics of fluids in general.

\begin{acknowledgements} 
We thank WestGrid for providing computational resources, and NSERC for financial support.
\end{acknowledgements}


%

\end{document}